# Quality indicators for scientific journals based on experts' opinion

Elea Giménez-Toledo(1), Jorge Mañana-Rodríguez(2), Emilio Delgado-López-Cózar(3)

**(1)Elea Giménez-Toledo** (corresponding author) has a PhD in information science. She is research fellow at the CSIC and director of EPUC, which is devoted to the evaluation of social science and humanities journals and books. She is co-author of two platforms on Spanish journal evaluation on social sciences and humanities: RESH and DICE (Dissemination and Editorial Quality of Spanish Journals in the Humanities, Social Sciences, and Law).Spanish National Research Council, Centre for Human and Social Sciences, Albasanz, 26-28, 28037, Madrid, Spain. E-mail: elea.gimenez@cchs.csic.es

**(2)Jorge Mañana-Rodríguez** is PhD student at the Spanish National Research Council, and member of EPUC research group. His work is devoted to the study of specialization measurement, characteristics and implications for scientific output assessment in social sciences and humanities. Spanish National Research Council, Centre for Human and Social Sciences, Albasanz, 26-28, 28037, Madrid, Spain. E-mail: jorge.mannana@cchs.csic.es. Telephone (0034) 916022795

**(3)Emilio Delgado-López-Cózar** is Professor of Library and Information Science at the University of Granada and member of EC3 research group. He is devoted to the evaluation of scientific journals and science, the study of research in LiS and to the evaluation of scientific output. He is one of the developers of In-Recs Index (Impact factor of Spanish Social Sciences Journals). Department of library and information science, Granada University, Cartuja Campus,18011, Granada, Spain. E-mail: edelgado@ugr.es

**Abstract:**

This paper presents the results and further development of a survey sent to 11,799 Spanish faculty members and researchers from various fields of the social sciences and the humanities, obtaining a total of 45.6% (5,368 responses) usable answers. Respondents were asked (a) to indicate the three most important journals in their field and (b) to rate them on a 0-10 scale according to their quality. The information obtained has been synthesized in two indicators which reflect the perceived quality of journals. Once the values were obtained, the journals were categorized according to each indicator and the ordinal positions were compared. Different profiles of journals are analyzed in connection with experts' opinion, such as regional orientation, and the consensus among researchers is studied. Finally, the possibilities of extending the research and indicators to sets of international journals are explored.

**Introduction**

Several quality indicators exist which are designed for the direct or indirect assessment of the quality of journals (Rousseau, 2002). Usually, a combination of these indicators is used by the most important databases in order to select journals (Thomson Reuters, 2012; Medline, 2012, for example), in the context of scientific assessment processes carried out by agencies or institutions, or by systems developed *ad hoc* to assess the quality of journals based in a given country.

Among these indicators it is possible to distinguish between bibliometric indicators (impact factor, internationality indicators, endogamy indicators, etc.), visibility indicators (i.e. presence in databases), indicators concerning editorial formal quality and the editorial management of publications (such as peer review and its characteristics). Nevertheless, although the combination of various indicators provides a more precise picture of the quality of a given journal, the humanities and social sciences scholarly community often claims that the assessment of the quality of content is a fundamental element in evaluation processes. The experts from each discipline are, in fact, the only ones who can provide this assessment of content. By its own nature, this assessment is subjective and biased, as the perception of "quality" is different for each person and can be influenced by their own concept of what rigor is in research, by their knowledge of the disciplinary scope of the journals they have to assess or by their involvement as an author or member of the advisory board in a given journal. In the case of the research conducted by Donohue and Fox (2000), using survey methodology and obtaining 243 usable answers, a positive but moderate correlation (r=0.58, n=46, p-value 0.0001) has been observed between the 5-year impact factor and the assessment made by the experts; in this sense, the moderate correlation shows a certain degree of discrepancy between these two measures.

Nevertheless, as has been proved in previous studies (Axarloglou & Theoharakis, 2003, Nederhoff & Zwaan, 1990), when the assessment is carried out by a large group of researchers in which various disciplines are represented, there is a convergence of opinions; that is to say, there is a concentration of votes by experts for a core of journals which could be considered key and as quality journals for each discipline. This could be particularly interesting in the assessment of the scientific activity processes in disciplines where the impact factor does not exist or is not a determinant. It is possible, as well, to identify journals which are considered not to be important, journals which are considered important by a limited number of faculty members and researchers, and journals which are considered important by most researchers. Above all, these assessments permit the calibration of the diversity of assessments made of different journals as a result of different schools of thought, approaches or areas of specialization. As Axarloglou & Theoharakis (2003) pointed out, diversity and pluralism in the disciplines does successfully contribute to their development and growth, but, nevertheless, Hodgson & Rothman (1999) pointed out that global research and the advisory boards of the top journals are dominated by a few institutions which defend their own ideas and approaches.

Assessing journals in a close-to-reality fashion involves taking into account the heterogeneity of the publications and opinions about these publications which can be found within a discipline, and this is one of the objectives of the survey on which this paper is based. The work of Axarloglou & Theoharakis (2003) is particularly illuminating in this sense, because it shows the differences in the assessment of the quality of journals provided by economists, depending on certain variables such as the school of thought in which these publications can be classified or the methodological approach followed in the research published in these journals. That is to say, they show that the perception of quality is neither coincident nor unanimous with the rankings or categorizations which are usually used in scientific assessment, and even varies among individuals belonging to an apparently homogeneous community, such as that of economists.

Nederhoff & Zwaan (1990) also addressed the measurement and analysis of perceived quality by Dutch and foreign scholars on a wide set of Dutch-based journals, using the survey as the information gathering technique. In the consultation of experts the respondents were asked to classify the journals as scholarly or non-scholarly, and a core of very important journals was identified which resulted from the overlapping of several quality characteristics.

The conclusions of the cited studies raise new questions regarding the debate between the practical "products" (databases, lists, etc.) which provide efficient solutions to scientific assessment processes and the more complex "products" which involve meeting the great diversity of parameters which affect the quality of publications. This idea is well expressed by Axarloglou & Theoharakis (2003, p.4) "The underlying perceptual heterogeneity with respect to Journal quality is a frequent cause of debate in tenure and promotion committees". In any assessment process, carefully carried out, that heterogeneity should be taken into consideration and, in this sense, a considerable number of indicators of a different nature could help to balance the weight and effects of a homogeneous assessment.

**Objectives and methodology**

It has been the general aim of this project to provide content quality indicators for Spanish social sciences and humanities journals. By doing this, the intention is to design and apply an extended range of quality indicators to scientific journals taken into account by the Spanish assessment agencies, which are available on websites such as DICE, RESH, In Recs, MIAR or CIRC, as examples of those most widely used. The main specific objective of this paper is to show and discuss the validity of the two quality indicators of scientific journals based on the opinion and content assessment provided by Spanish experts in the various disciplines of the humanities and the social sciences. For this purpose, the application of both indicators to the journals of two disciplines, anthropology and library and information science, is shown and the results and differences between both values are analyzed.

The information gathering instrument used is a web-form survey designed with PHP and MySQL, and the target population was comprised of 11,799 Spanish faculty members and researchers from various disciplines of the social sciences and the humanities, all which meet the condition of having, at least, a six year research period approved by the CNEAI[1].

Compared to other international and national surveys, this is one of the biggest samples of researchers that has been used up to now. Although personal invitations to participate in the survey were made, the answers were absolutely anonymous. This survey sought to identify the quality of the journals according to the opinion of the experts themselves and, at the same time, to observe the diversity among the disciplines belonging to the social sciences and the humanities. In this sense, another relevant feature of the survey is the level of data aggregation: scientific specialization and not disciplines; this is an important point since several studies have shown that scholars' specialization is a determinant in the perception of a journal's quality. In order to achieve this, it was important to distinguish the answers according to the discipline of the respondents.

---

[1] Recognition to the researchers in the form of additional payment which is given as a result of a positive assessment of six years of research activity.

The survey included a total of five questions regarding different aspects related to the quality of scientific journals, and two of the answers have been the basis for the design and development of the indicators proposed in this work:

*- Indicate the three best Spanish journals from your area of specialization and rate each of them from 1 to 10, being 1 the lowest value and 10 the highest value.*
*- Indicate the three best journals in your area of specialization. You can choose both Spanish and foreign journals. In order to rate each of them, consider 1 the lowest value and 10 the highest value.*

In order to facilitate the completion of these two questions, lists of journals by discipline or sets of disciplines were offered. These lists included more than 1,900 Spanish journal titles and more than 8,000 foreign journals (so, in the near future a comparison of perceptions between Spanish and foreign journals will be possible). If the respondent did not find the journal he wanted to select in these extensive lists, the respondent had the option of adding the title of the publication.

The response rate was 45.6% (5,368 answers) although it varied considerably according to the different disciplines of the respondents. This response rate is one of the highest among those found in similar studies (Lowe, 2005: 1,314 surveys sent, 149 usable answers, 16% response rate; Axarglolou, 2003: 10,402 mails sent, 2,103 usable answers,: 20.22% response rate; Brinn, 1996: 260 surveys sent, 90 usable answers, 34.6% response rate; Giles, 1989: 550 surveys sent, 215 usable answers, 40% response rate), although it does not reach the 69.4% of Nederhoff & Zwaan (1990) who, however, had a more limited target set of researchers (385 surveys). This high response rate can be explained by the strong

"awareness" of the community of Spanish scholars and teachers in the disciplines of the humanities and the social sciences regarding scientific assessment issues.

The survey provides two kinds of variables from which the indicators have been developed. Firstly, the number of votes or times a given journal has been rated in any position (called by Axarloglou "familiarity" in his study). Secondly, the position (first, second or third) in which a given journal was voted. This indicator is related to the Average Rank Position which was proposed by Hull & Wright (1990) and also mentioned in the cited study, whereas the number of votes would be related to the concept of familiarity as detailed in Theoharakis & Hirst (2002) as well as in Oltheten (2005). In the case of that study, familiarity was described as the number of times a given journal was positioned in the upper 20% of top quality journals.

**Quality indicators according to experts**

The information obtained from the survey is presented in a table similar to the following, which will be used as an example for the explanation and calculus of the two indicators developed.

**Table 1: Example of the data obtained from the survey**

| TITLE | NUMBER OF TIMES VOTED 1ST MOST IMPORTANT JOURNAL | SCORES WHEN VOTED 1ST MOST IMPORTANT JOURNAL | SUM OF SCORES WHEN VOTED IN 1ST POSITION | NUMBER OF TIMES VOTED 2ND MOST IMPORTANT JOURNAL | SCORES WHEN VOTED 2ND MOST IMPORTANT JOURNAL | SUM OF SCORES WHEN VOTED IN 2ND POSITION | NUMBER OF TIMES VOTED 3RD MOST IMPORTANT JOURNAL | SCORES WHEN VOTED 3RD MOST IMPORTANT JOURNAL | SUM OF SCORES WHEN VOTED IN 3RD POSITION |
|---|---|---|---|---|---|---|---|---|---|
| A | 3 | 7; 8; 6 | 21 | 4 | 6; 7; 9; 5 | 27 | 2 | 6; 5 | 11 |
| B | 6 | 8; 7; 7; 9; 6; 8 | 45 | 2 | 6; 8 | 14 | 1 | 6 | 6 |
| C | 2 | 9; 7 | 16 | 6 | 7;5;6;8;5;6 | 37 | 4 | 5;4;7;6 | 22 |
| ∑ | 11 | | 82 | 12 | | 78 | 7 | | 39 |

From the information given above a ranking can be derived. The information related to the quality of a journal is both the number of votes it has received as the first, second or third most important journal, as well as the scores the journal has received when voted in each position.

When the scores received by a journal as the first most important journal (or in the other two positions, second or third) are added together, the frequency with which the journal has been voted affects the overall sum of the scores; an increase of 1 in the number of votes means an increase of between 1 and 10 in the sum of scores for that position (as the first, second or third most important journal). A journal could be voted three times as the first most important and the scores could be 5, 5, and 5 for each vote, while another could also be voted three times as the first most important journal but receive a score of 10, 10 and 10. An additional vote for a given position (first, second or third) will always mean an increase in the sum of scores for a journal. From this direct relation between the number of votes and the sum of scores, it can be derived that the sum of the scores given to a journal is a representative measure of its perceived quality. Nevertheless, a single journal can be voted a different number of times (with different associated scores) in different positions. The values of the scores given to a journal can vary from 1 to 10, regardless of the position that journal is being voted for, and at the same time the indicator should be sensitive not only to the sum of scores, but also to the different positions (first, second or third) among which this scores are distributed.

**Table 2: Journals with the same overall score, but received in different positions.**

The sum of the scores, in first, second and third position, for journals M and N is the same: 43, but the score given to M in the first position is 1 higher than journal N, whereas N scores 1 higher when voted as the third most important journal. This makes the use of a weight necessary.

The general formula of the indicator would be the following:

| TITLE | NUMBER OF TIMES VOTED 1ST MOST IMPORTANT JOURNAL | SCORES WHEN VOTED 1ST MOST IMPORTANT JOURNAL | SUM OF SCORES WHEN VOTED IN 1ST POSITION | NUMBER OF TIMES VOTED 2ND MOST IMPORTANT JOURNAL | SCORES WHEN VOTED 2ND MOST IMPORTANT JOURNAL | SUM OF SCORES WHEN VOTED IN 2ND POSITION | NUMBER OF TIMES VOTED 3RD MOST IMPORTANT JOURNAL | SCORES WHEN VOTED 3RD MOST IMPORTANT JOURNAL | SUM OF SCORES WHEN VOTED IN 3RD POSITION |
|---|---|---|---|---|---|---|---|---|---|
| M | 2 | 9 ;10 | 19 | 2 | 7;7 | 14 | 2 | 5;5 | 10 |
| N | 2 | 9; 9 | 18 | 2 | 7;7 | 14 | 2 | 5;6 | 11 |

$$V = \sum_{i=f}^{i=t} \Sigma S_i * W_i = \Sigma S_f * W_f + \Sigma S_s * W_s + \Sigma S_t * W_t$$

Where:

$i$ can take three values: $f$ denotes first position, $s$ denotes the second position and $t$ denotes the third position.

$S_f$: Are the scores given when voted as the first most important journal.

$S_s$: Are the scores given when voted as the second most important journal.

$S_t$: Are the scores given when voted as the third most important journal.

$W_f$: Is the weight applied to the sum of scores when voted as the first most important journal.

$W_s$: Is the weight applied to the sum of scores when voted as the second most important journal.

$W_t$: Is the weight applied to the sum of scores when voted as the third most important journal.

Regarding the value of the weight, the only condition it has to meet is: $W_f > W_s > W_t$. This is because the scores given to a journal when voted as the first most important should be weighted with a higher value than when that same journal has been voted as the second or the third most important.

To ensure that this condition is met, two options are proposed. Firstly, it is possible to give an arbitrary weight to each sum of scores for first, second and third position. As an example, the value of $W_f$ could be 3, the value of $W_s$, 2 and the value of $W_t$. This indicator is denominated $V_1$:

$$V1 = \sum_{i=f}^{i=t} \Sigma S_i * W_{i1} = \Sigma S_f * 3 + \Sigma S_s * 2 + \Sigma S_t$$

The generic weight $Wi$ is notated here as $Wi1$ in order to distinguish it from the weight applied to the second indicator $V_2$ (explained below) where the weight is noted as $Wi2$.

From table 1, indicator $V_1$ can be calculated, i.e. for journal A as follows:

**$V_1$= 21\*3+27\*2+11= 128**

Nevertheless, to assign a weight taking as the only condition that the values meet the mentioned condition could raise these questions: is it appropriate to give a weight to the first position which is three times larger than the weight given to the third position? Furthermore, it should be considered whether assigning an arbitrary and equal weight to all journals regardless of their discipline is the most adequate option, taking into account that the differences between the average scores given to journals voted as the first, second and third most important journals substantially vary among different disciplines and areas of specialization.

As a solution to these questions, it is proposed to deduce the value of the weights from the information given by the respondents itself, that is to say, from the results of the survey. By doing this, a possible value for the weight would be the average of the scores given by the respondents for each position (first, second or third); if the respondents give a high average score to the journals in first position, a smaller average score in the second position and an even smaller average in the third position, the condition for the weights would be met. Moreover, this weight would not be arbitrary, because it would be adjusted to the values given by the respondents for each discipline.

The weight proposed for indicator $V_2$ involves two measures for its value in each position, first, second or third.

The **first** measure is the average score per vote (ASV) in each position for all journals: the quotient of the total score given to all journals and the total number of votes received by all journals in each position.

In the case of journals in table 1:

$$ASV_1 = \frac{82}{11} = 7.45; \quad ASV_2 = \frac{78}{12} = 6.5; \quad ASV_3 = \frac{39}{7} = 5.57$$

For all disciplines it has been observed that $ASV_1 > ASV_2 > ASV_3$, which means that these average scores meet, in all cases, the condition required. It is sensitive to differences in the scoring pattern, which differs among the different disciplines.

Nevertheless, these values are too big to be suitable weights: the values potentially range from 1 to 10, which in some cases is more than the sum of the values given by the respondents in the 1-10 scale. A suitable solution to this problem is to include, as a denominator, the **second** measure used to develop this second weight, which is the sum of the three ASV. It is the same as expressing the average score per vote for each position in a per-unit range with respect to the sum of the three averages. This is $W_{t2}$, the weight applied in the case of the second indicator $V_2$.

A further advantage of using this per-unit average score per vote is that the sum of the three values for the weights would always be 1. On the contrary, if "raw" ASV values where used as weights for the sum of scores given in each position, their sum would not be the same among different disciplines, which could involve comparability problems. The general formula for this weight is:

$$W_{t2} = \frac{ASV_k}{\sum_{i=f}^{i=t} ASV_i}$$

Using the data in table 1, the values of the weights applied to the sum of scores in first, second and third position would be:

$$W_{f2} = \frac{7.45}{7.45 + 6.5 + 5.57} = 0.38$$

$$W_{s2} = \frac{6.5}{7.45+6.5+5.57} = 0.33$$

$$W_{t2} = \frac{5.57}{7.45+6.5+5.57} = 0.28$$

Finally, the value of indicator $V_2$ would be:

JOURNAL A: $V_2 = 21* 0.38 +27 *0.33+11 *0.28=$ **19.97**

JOURNAL B: $V_2 = 45* 0.38 +14 *0.33+6 *0.28=$ **23.4**

JOURNAL C: $V_2 = 16* 0.38 +37 *0.33+22 *0.28=$ **24. 45**

**Results**

The design of the survey itself permits two variables to be obtained for each journal: the number of votes and the scores given to the journal, and the position in which that journal has been voted. Moreover, taking into account that each discipline has a different number of journals, a different population of scholars and teachers who could potentially give a score for each of the journals, and a different appreciation of the journals (the degree of usage is different among the disciplines comprising the humanities and the social sciences), it seemed logical that the indicator for the publications should take into account the behavior of the discipline in the survey, that is to say, it should not be calculated on all the journals from the opinion of all the respondents. Therefore, both proposed formulae are related to the answer obtained in the context of a particular discipline. Within a discipline there are different "knowledge areas" (areas of specialization), recognized by the former Innovation and Science Ministry (Spain), for which the votes, positions and scores have been clustered according to this disciplinary structure in order to give the value of the indicator for each discipline.

Both indicators have been applied to the same set of journals, and the variations observed in the change of position for the same journal between the two indicators seem to be strictly related to the values of the weights W. In the case of the second indicator ($V_2$), when the weights, which are dependent on the distribution of scores, are close to 3, 2 and 1 for the first, second and third positions respectively, both indicators would offer almost identical ordinal values. As an example of this, in the following table the values and position changes between both indicators for Spanish Anthropology journals are shown:

**Table 3. Spanish Anthropology journals. $V_1$ vs. $V_2$: values and position comparison.**

| TITLE | $V_1$ | Position change from $V_1$ to $V_2$ | $V_2$ |
|---|---|---|---|
| Revista de Antropología Social | 52,3 | 0 | 24,77 |
| Revista de Dialectología y Tradiciones Populares | 21 | 0 | 15,76 |
| Revista d'Etnologia de Catalunya | 7,67 | 0 | 9,51 |
| AIBR. Revista de Antropología Iberoamerica | 5,56 | 0 | 8,41 |
| Historia, Antropología y Fuentes Orales | 2,08 | 0 | 5,56 |
| Gazeta de Antropología | 1,72 | 0 | 4,75 |
| Trans. Revista Transcultural de Música | 1,62 | 0 | 4,63 |
| Ankulegi. Revista de Antropología Social | 0,87 | 0 | 3 |
| Pasos. Revista de Turismo y Patrimonio Cultural | 0,46 | 0 | 2,36 |
| Demófilo. Revista de Cultura Tradicional de Andalucía | 0,31 | 0 | 1,66 |
| Revista de Antropologia Experimental | 0,14 | 0 | 1,43 |
| Anales de la Fundación Joaquín Costa | 0,12 | 0 | 1,29 |
| Oráfrica | 0,05 | 0 | 1,1 |
| Anuario de Eusko-Folklore | 0,02 | 0 | 0,4 |
| Cuadernos de Etnología y Etnografía de Navarra | 0,01 | 0 | 0,37 |
| Culturas Populares | 0,01 | 0 | 0,37 |
| Etniker Bizkaia | 0,01 | 0 | 0,37 |
| Revista Valenciana d'etnologia | 0,01 | 0 | 0,33 |
| Temas de Antropología Aragonesa | 0,01 | 0 | 0,31 |

Nevertheless, there are position changes directly proportional (among other factors related to the calculus of the indicator) to the magnitude of the differences between the weights used in the indicators $V_1$ y $V_2$. As an example of this characteristic, in the following table, shown are the values of both indicators and the position changes in the ordinal scale for the Information & Library Science discipline.

**Table 4. Spanish Library & Information Science journals. $V_1$ vs. $V_2$: values and position comparison.**

| TITLE | $V_1$ | Position change from $V_1$ to $V_2$ | TITLE | $V_2$ |
|---|---|---|---|---|
| El Profesional de la Información | 40,15 | -1 | Revista Española de Documentación Científica | 26,80 |
| Revista Española de Documentación Científica | 31,31 | 1 | El Profesional de la Información | 25,62 |
| Lligall. Revista Catalana d'Arxivística | 6,53 | -6 | BiD: textos universitaris de biblioteconomia i documentació | 8,61 |
| Revista General de Información y Documentación | 4,41 | -4 | Scire. Representación y Organización del Conocimiento | 6,34 |
| BiD: textos universitaris de biblioteconomia i documentació | 4,04 | 3 | Anales de Documentación | 6,21 |
| Anales de Documentación | 2,69 | 1 | Boletín de la ANABAD | 5,14 |
| Scire. Representación y Organización del Conocimiento | 2,14 | 3 | Cybermetrics. International Journal of Scientometrics, Informetrics and Bibliometrics | 4,59 |
| Documentación de las Ciencias de la Información | 2,05 | -2 | Revista General de Información y Documentación | 4,30 |
| Boletín de la ANABAD | 1,69 | 3 | Lligall. Revista Catalana d'Arxivística | 3,40 |
| Cybermetrics. International Journal of Scientometrics, Informetrics and Bibliometrics | 0,83 | 3 | Documentación de las Ciencias de la Información | 2,96 |
| Item. Revista de Biblioteconomía i Documentació | 0,13 | 0 | Item. Revista de Biblioteconomía i Documentació | 1,73 |
| Ibersid. Revista de Sistemas de Información y Documentación | 0,05 | 0 | Ibersid. Revista de Sistemas de Información y Documentación | 1,26 |
| Cuadernos de Documentación Multimedia (ed. electrónica) | 0,02 | 0 | Cuadernos de Documentación Multimedia (ed. electrónica) | 0,75 |
| Revista de Museología | 0,02 | 0 | Revista de Museología | 0,27 |

The tabulation of the results of the survey has offered the possibility; furthermore, of detecting multidisciplinary journals mentioned by respondents and which belong to different knowledge fields. A given journal can be included in various disciplines and obtain different values in each of these disciplines. This aspect opens the door to research in the context of multidisciplinary journals, referring to the consequences which the evaluation of journals could have for publications which have a multidisciplinary scope and which, maybe due to this factor, are not recognized as core publications in any discipline; that is to say, the possible penalization of specialized or multidisciplinary journals in assessment systems.

**Discussion and conclusions**

The assessment of a journal's content through expert opinion should be one more variable to be considered in the assessment of scientific journals. By doing so, formal or indirect quality indicators would be complemented, and the scientific knowledge of scholars and the value given by them to the journals in their disciplines would be taken into account and ascertained.

This proposal is not only based on the votes obtained by a journal given by the respondents, but also on the score and position given to it. In this sense, it differs from the method used by Nederhoff & Zwaan (1990), based on average scores received by the journals. The indicators proposed in this study show more points in common with those presented by Axarloglou & Theoharakis (2003). Basically, in all the formulae shown above, taken into account is the frequency with which a journal has been voted, the position assigned to it and a specific weight which allows both the generation of a ranking and evidence of the differences among journals. In the case of the study by Axarloglou & Theoharakis there is the variable "tier" in which respondents positioned the journal; respondents were asked to position each of the voted journals among the first 15 journals (first tier) or among the second 15 (second tier). This difference is fundamental, because the weight given to each journal depends on the tier in which it is positioned. The weight given by Axarloglou & Theoharakis to each journal is a constant value and the difference between the positions would be always a thirtieth, since the number of journals in both tiers is 30. In this paper, the weight is given by the average of the scores and the position in which the journals of each discipline appear (first, second or third position).

This represents an important advantage in the assessment of scientific activity in the humanities and the social sciences where different schools of thought, different positioning depending on the methodology used and a wide range of areas of specialization with different degrees of international projection can be found. To have an assessment of the content of the journals, which takes into account the characteristics of the sample, allows the particularities of each discipline to be attended to, but it does not allow raw comparisons between disciplines.

One of the results of Axarloglou & Theoharakis is that, although there is agreement among respondents regarding the top journals, it is also true that larger differences can be observed in relation to the more regional oriented journals. The geographic factor, the school of thought, the participation in or affiliation to a journal, the research methodology approach or the area of specialization and the characteristics of the institution of the respondent could affect and cause variations in the perception of the quality of a journal. All these factors allow them to affirm that "These findings should serve as a warning against monolithic research evaluation practices that do not account for the underlying differences of the community" (Axarloglou & Theoharakis, 2003). In this sense, recent developments in Flanders (Engels et al, 2012) take into account a variety of sources and assessment procedures, among them is expert opinion in the form of panels, though survey methodology is not used.

In the formulae of both indicators $V_1$ and $V_2$ a greater preciseness can be appreciated in $V_2$. Mathematically, the second indicator ($V_2$) could be considered more adjusted to the distribution of votes in different disciplines. If the value of the indicator is to be for public use, and therefore is going to be a reference and data to be taken into account by different specialists (evaluators, researchers, editors), it is recommended that the formulae be the easiest and the most precise possible. Too complex an indicator could cause mistrust among users and, in some cases, make comprehension difficult.

The results of the application of these indicators reveal the extent to which they are comparable and if there is a parallel or not between the values. The results conclude that there is a high correlation between them, although small differences are observed in the position of certain journals. A line of research is now open to analyze the values of $V_2$ for all journals, so that the opinion of specialists regarding the journals in their disciplines will become better known and described in more detail.

Nevertheless, the team which has developed this study decided to include the $V_1$ indicator in the RESH[2] information system which makes a comprehensive assessment of all the Spanish scholarly journals in the humanities and the social sciences. The diversity of indicators in this information system will allow the correlation of variables such as the impact of journals and the assessment provided by experts. Nederhoff & Zwaan (1990) have worked previously on this last point, finding as one of the results that the journals with the best expert assessment score did coincide with those with the highest impact factor, but also finding cases in which the tendency is exactly the opposite.

The indicators shown in RESH related to the presence of journals in international databases will give also the opportunity of identifying if the publications which have the best score according to expert opinion are also those with greater dissemination or have been selected by the international databases such as WoS or Scopus. In this sense, it would be interesting to observe whether the journals with the best scores for journals belonging to the disciplines with a strong local orientation are covered or not by these databases and, therefore, probably certain errors might be identified, which often occur in the assessment processes of the output of the humanities and the social sciences when these databases are the only source of information taken into account.

---

[2] Epuc.cchs.csic.es/resh

Although expert consultation seems to be *a priori* the best method for making an approximated assessment of the content quality of journals, such an assessment should be handled carefully due to various reasons. Firstly, the assessment by an expert is, in fact, a perception of quality. This perception is necessarily influenced by the expert's academic life: reading research papers, the degree of participation on advisory or editorial boards, or even the fact of publishing in journals which have assessment processes of papers, as well as the invisible colleges which exist in the various disciplines or sub-disciplines of journals. The journal is, like it or not, a part of a network or an invisible college.

In the case of Spain, it has taken time, effort and many research projects to meet the needs of the social sciences and the humanities, evaluating journals with different methods and adapting these to the characteristics of particular disciplines. There is still much to be done and one of the steps to be fulfilled would be to deal with the differences and diversity of publishing habits within a given discipline which is reflected in the publications selected by scholars. Even in the discipline of Economics, a "hard science" in the context of the social sciences, differences are observed and argued.

**Funding and acknowledgements**

This study has been carried out within the framework of the research project *Valoración integrada de las revistas españolas de Ciencias Sociales y Humanidades mediante la aplicación de indicadores múltiples* [Integrated Assessment of Spanish Social Sciences and Humanities journals by the application of multiple indicators] SEJ2007-68069-C02-02, funded by the Spanish Ministry of Science and Innovation between 2007 and 2010.

The authors would like to acknowledge Sonia Jiménez, the Information and Communication Technologies and Statistical Analysis units for their contribution to this work.

DICE: epuc.cchs.csic.es/dice

RESH: epuc.cchs.csic.es/resh

CIRC: epuc.cchs.csic.es/circ

MIAR: http://miar.ub.es/que.php

In Recs: http://ec3.ugr.es/in-recs